\documentclass[12pt]{article}
\usepackage[dvips]{graphicx,psfrag}
\usepackage{amssymb}
\input epsf
\newcommand{\be}{\begin{equation}}
\newcommand{\ee}{\end{equation}}
\newcommand{\bea}{\begin{eqnarray}}
\newcommand{\eea}{\end{eqnarray}}
\newcommand{\lb}{\label}
\begin{document}
\begin{titlepage}
\begin{center}
{\large\bf 
GALACTIC COSMIC RAYS FROM PBHs AND PRIMORDIAL SPECTRA WITH A SCALE}
\vskip 2 cm
{\bf Aur\'elien Barrau, David Blais$^*$, Ga\"elle Boudoul, David Polarski$^*$}
\vskip 0.4cm
 Institut des Sciences Nucl\'eaires de Grenoble UMR 5821 CNRS-IN2P3,\\
 Universit\'e Joseph Fourier, Grenoble-I, France.
\vskip 0.7cm
 $^*$ Laboratoire de Physique Math\'ematique et Th\'eorique,
 UMR 5825 CNRS,\\
 Universit\'e de Montpellier II, 34095 Montpellier, France.

\end{center}

\vskip 1cm
\small
\begin{center}
{\bf Abstract}
\end{center}

\begin{quote}
We consider the observational constraints from the detection of antiprotons 
in the Galaxy on the amount of Primordial Black Holes (PBH) produced from 
primordial power spectra with a bumpy mass variance. Though essentially 
equivalent at the present time to the constraints from the diffuse 
$\gamma$-ray background, they allow a widely independent approach and they 
should improve sensibly in the nearby future. We discuss the resulting 
constraints on inflationary parameters using a Broken Scale Invariance (BSI) 
model as a concrete example.
\end{quote} 
\normalsize
\date{\today}

PACS Numbers: 04.62.+v, 98.80.Cq
\end{titlepage}

\section{Introduction}
The formation of PBHs in the early universe is an inevitable prediction based on 
general relativity, the existence of a hot phase and, most importantly, the 
presence of primordial fluctuations which are the seed of the large structures 
in our universe \cite{ZN67}. It can have many interesting cosmological 
consequences and is one of the few constraints available 
on the primordial fluctuations on very small scales that can be based on existent 
astrophysical observations (see e.g. \cite{C85}). 
It has been used by various authors in order to constrain the 
spectrum of primordial fluctuations, in particular in order to find an upper 
limit on the spectral index $n$ and on the present relative density of PBHs 
with $M\approx M_*$ (the initial mass of a PBH whose 
lifetime equals the age of the Universe) \cite{CL93},\cite{GL97},\cite{BBKP02}.
A possible contribution of evaporating PBHs to the diffuse $\gamma$-ray background 
is presently the most constraining observation \cite{CM98}. 
On the other hand, the observation of antiprotons in the Galaxy \cite{McGC91} is as 
powerful \cite{M96} and, in contrast to the $\gamma$-ray background, sensitive improvements 
can be expected in the near future. These involve both experimental and 
theoretical progress. This is why it is interesting to consider in some details 
the constraints these observations can, and will, put on any primordial fluctuations 
model, and prominently on some inflationary models. As noted earlier (see, e.g., 
Fig.1 in \cite{BKP02a}), a constant 
spectral index $n$ would need extreme fine tuning in order to saturate the 
$\gamma$-ray or antiproton constraint, and such a large $n$ is anyway excluded 
by the latest CMB data. Hence we consider here spectra with a characteristic scale 
for which the generation of PBHs is boosted in a certain mass range.     
\section{PBH formation and primordial fluctuations}
\par\noindent
{\it Density of PBHs from bumpy mass variance}: 
For detailed confrontation with cosmological and astrophysical 
observations one often needs the mass spectrum, the number density 
per unit of mass. This is particularly delicate for PBHs and we follow 
here a derivation valid in the presence of a bump, as given in \cite{Y98}.
The first assumption is that the primordial spectrum of cosmological 
fluctuations has a characteristic scale in its power spectrum $P(k)$, 
which results in a well-localized bump in its mass variance. The importance 
of this assumption lies in the determination of the PBH mass scale $M_{peak}$ 
where PBH formation mainly occurs.  
The second assumption, supported by numerical simulations, is that PBH 
formation occurs through near-critical collapse \cite{NJ98} whereby PBH 
with different masses $M$ around $M_{peak}\equiv M_H(t_{k_{peak}})$, the 
horizon mass at the (horizon-crossing) time $t_{k_{peak}}$ -- the horizon 
crossing time $t_k$ is defined through $k=a(t_k) H(t_k)$ -- could be formed 
at the same time $t_{k_{peak}}$, according to 
\be
M = \kappa~M_H ( \delta - \delta_c )^{\gamma}~,\lb{cc}
\ee
where $\delta_c$ is a control parameter.
While the parameters $\gamma$ and $\delta_c$ are universal with 
$\gamma\approx 0.35$, $\delta_c\approx 0.7$, the parameter $\kappa$ 
(or $\epsilon$, see below) can vary sensibly and fixes essentially 
the typical PBH mass. As shown in \cite{Y98}, one finds 
\be
\frac{d\Omega_{PBH}}{d\ln M}\equiv \frac{d \Omega_{PBH}(M,t_{k_{peak}})}{d\ln M} 
    = \left( \gamma~\kappa^{\frac{1}{\gamma}}\right )^{-1}~
\left( \frac{M}{M_{peak}}\right )^{1+\frac{1}{\gamma}} p[\delta(M)].\lb{mf1}
\ee
If we identify the maximum of (\ref{mf1}) in the following way
\be
M_{max} = \epsilon~M_{peak}~,
\ee
we are led to the result
\be
\frac{d \Omega_{PBH}}{d\ln M} = \epsilon^{-\frac{1}{\gamma}}~\beta(M_{peak})
      ~(1 + \frac{1}{\gamma})~\left ( \frac{M}{M_{peak}}\right )^{1+ \frac{1}{\gamma}}
        \exp \Biggl [-\epsilon^{-\frac{1}{\gamma}} (1+\gamma) 
        \left ( \frac{M}{M_{peak}}\right )^{\frac{1}{\gamma}}\Biggr ]~,\lb{mf2}    
\ee     
$\beta(M_{peak})$ gives the probability that a region 
of comoving size $R=(H^{-1}/a)|_{t=t_{k_{peak}}}$ has an averaged 
density contrast at the time $t_{k_{peak}}$ in the 
range $\delta_c\leq\delta\leq\delta_{max}$ 
\be
\label{beta1}
  \beta(M_{peak})= \int_{\delta_c}^{\delta_{max}} 
                                p(\delta, t_{k_{peak}})~\textrm{d}\delta~.
\ee
It is then straightforward to find the quantity of interest to us 
\be
\frac{d^2 n_i}{dM_i~dV_i} =\frac{3M^2_p}{32\pi}~\left(\frac{M_p}{M_{peak}}\right)^4 
           x^{-2} ~\frac{d \Omega_{PBH}}{d\ln M}(x)~,\lb{mf3}
\ee
where $M_p$ stands for the Planck mass while $x\equiv \frac{M}{M_{peak}}$.
The subscript $i$ stands for ``initial'', i.e. at the time of formation.
The mass $M_{peak}$ 
corresponds to the maximum in the mass variance $\sigma_H(t_k)$ and 
{\it not} to the maximum in the primordial spectrum itself \cite{BBKP02}. 
The parameters $\gamma$ and $\epsilon$ refer to PBH formation while 
$M_{peak}$ and $\beta(M_{peak})$ refer to the primordial spectrum.
\vskip 20pt
\par\noindent
{\it Primordial inflationary fluctuations}:
One usually considers Gaussian primordial inflationary fluctuations but 
it should be stressed that non-Gaussianity of the fluctuations could 
lead to sensibly different results \cite{BP97}. For primordial fluctuations 
with a Gaussian probability density $p[\delta]$, we have 
\be
p(\delta) = \frac{1}{\sqrt{2\pi}~\sigma (R)}~ 
e^{-\frac{\delta^2}{2 \sigma^2(R)}}~,~~~~~~~~~~~\sigma^2(R) = \frac{1}{2\pi^2}
        \int_0^{\infty}dk ~k^2 ~W^2_{TH}(kR) ~P(k)\lb{sigW}~,
\ee 
where $\delta$ is the density contrast averaged over a sphere of radius $R$, and 
$\sigma^2(R) \equiv \Bigl \langle \Bigl ( \frac{\delta M}{M}  
\Bigr )_R^2 \Bigr \rangle$ is computed using a top-hat window function. 
%
%
Usually what is meant by the primordial power spectrum is the
power spectrum on superhorizon scales after the end of inflation. 
On these scales, the scale dependence of 
the power spectrum is unaffected by cosmic evolution. On subhorizon 
scales, however, this is no longer the case, and 
%
%
one has instead
\be\lb{T}
P(k,t)= \frac{P(0,t)}{P(0,t_i)}~P(k,t_i)~T^2(k,t)\; , \quad T(k\to 0,t)\to 1~,
\ee
where $t_i$ is some initial time when all scales are outside the Hubble radius 
($k<aH$). Therefore, the power spectrum 
$P(k)$ on sub-horizon scales appearing in (\ref{sigW}) must involve 
convolution with the transfer function {\it at time} $t_k$ \cite{BKP02a}.
At reentrance inside the Hubble radius during the radiation dominated stage, one has 
{\it in complete generality} \cite{P02},\cite{BBKP02} 
(subscript $e$ stands for the end of inflation)
\be
\sigma_H^2(t_{k})=\frac{8}{81\pi^2} \int_0^{\frac{k_e}{k}} x^{3}~F(kx)~T^2(kx,t_k)~
                              W^2_{TH}(x)~\textrm{d}x~,~~~~~~~~~~t_{k_e}\ll t_k\ll t_{eq}~,
\lb{sigF} 
\ee
where the transfer function can be computed analytically and yields   
\be
T^2(kx,t_k)\equiv \Biggl[\frac{9}{x^2}\left(\frac{\sin(c_s x)}{c_s x}-\cos(c_s x)\right) 
\Biggr]^2= W^2_{TH}(c_sx)= W^2_{TH}\left(\frac{x}{\sqrt{3}}\right)~,\lb{TW}
\ee 
while $F(k)\equiv \frac{81}{16} k^3 P(k,t_k)=\frac{81}{8} \pi^2 \delta^2_H(k,t_k)$. 
Finally $\beta(M_{peak})$ is given by
\be
\label{beta}
  \beta(M_{peak})\approx \frac{\sigma_H(t_{k_{peak}})}{\sqrt{2\pi}\,\delta_c}
           e^{-\frac{\delta_c^2}{2 \sigma_H^2(t_{k_{peak}})}}~,
\ee
with $\sigma_H^2(t_{k_{peak}})\equiv \sigma^2(R)\big|_{t_{k_{peak}}}\equiv \sigma^2(M_{peak})$, 
and we will take $\delta_c=0.7$.

For a given primordial fluctuations spectrum of inflationary origin normalized at 
large scales using the COBE data, the quantities $M_{peak}$ and $\beta(M_{peak})$ 
can be computed numerically and will depend on some inflationary parameters 
specifying that model as well as on cosmological parameters pertaining to the 
cosmological background evolution \cite{P02}. On the other hand $\gamma$ and 
$\epsilon$ should be found by numerical simulations of PBH formation for this 
particular spectrum. 
Values $\epsilon=0.5,~1,~2$, correspond to $\kappa\approx 2.7,~5.4,~10.8$.

\section{Evaporation, fragmentation and source term}

As shown by Hawking \cite{Hawking5}, such PBHs should evaporate into
particles of energy $Q$ per unit of time $t$ (for each degree of
freedom):
\begin{equation}
\frac{{\rm d}^2N}{{\rm d}Q{\rm
d}t}=\frac{\Gamma_s}{h\left(\exp\left(\frac{Q}{h\kappa/4\pi^2c}\right)-(-1)^{2s}\right)}~,
\end{equation}
where contributions of angular velocity and electric potential have
been neglected since the
black hole discharges and finishes its rotation much faster than it
evaporates \cite{Gibbons}. The quantity $\kappa$ is the surface 
gravity, $s$ is the
spin of the emitted species and
$\Gamma_s$ is the absorption probability. If the Hawking
temperature, defined by 
$
T=hc^3/(16\pi k G M)\approx (10^{13}{\rm g}/M)~{\rm GeV}
$ is introduced,
the argument of the exponent becomes simply a function of $Q/kT$.
Although the absorption probability is often approximated by its
relativistic limit,
we took into account in this work its real expression for
non-relativistic particles:
\begin{equation}
\Gamma_s=\frac{4\pi \sigma_s(Q,M,\mu)}{h^2c^2}(Q^2-\mu^2)~,
\end{equation}
where $\sigma_s$ is the absorption cross section computed numerically
\cite{Page2} and $\mu$ is the rest mass of the emitted particle.

Among other cosmic rays emitted by evaporating PBHs, antiprotons are 
especially interesting as
their secondary flux is both rather small
(the $\bar{p}/p$ ratio near the Earth is lower than $10^{-4}$ at all energies) and
quite well known \cite{fio}. We will therefore focus on such antiparticles in this paper.
As shown by MacGibbon and Webber \cite{MacGibbon1}, when the
black hole temperature is
greater than the quantum chromodynamics confinement scale
$\Lambda_{QCD}$, quarks and gluons jets are
emitted instead of composite hadrons. To evaluate the number of
emitted antiprotons $\bar{p}$ , one therefore
needs to perform the following convolution:
\begin{equation}
\frac{{\rm d}^2N_{\bar{p}}}{{\rm d}E{\rm d}t}=
\sum_j\int_{Q=E}^{\infty}\alpha_j\frac{\Gamma_{s_j}(Q,T)}{h}
\left(e^{\frac{Q}{kT}}-(-1)^{2s_j}\right)^{-1}
\times\frac{{\rm d}g_{j\bar{p}}(Q,E)}{{\rm d}E}{\rm d}Q~,
\end{equation}
where $\alpha_j$ is the number of degrees of freedom, $E$ is the
antiproton energy and
${\rm d}g_{j\bar{p}}(Q,E)/{\rm d}E$ is the normalized differential
fragmentation function, {\it i.e.}
the number of antiprotons between $E$ and $E+{\rm d}E$ created by a
parton jet of type $j$ and energy
$Q$. The fragmentation functions have been evaluated with the
high-energy physics event generator
PYTHIA/JETSET \cite{Tj} based on the string fragmentation model.

Once the spectrum of emitted antiprotons is known for a single PBH of given mass, the
source term used for propagation can be obtained through
\begin{equation}
\frac{d^3 N_{\bar{p}}^{\odot}}{dEdtdV}(E) = \int_{0}^{\infty}
\frac{d^2N_{\bar{p}}}{dEdt}(M,t_0)\frac{d^2n}{dMdV_i}dM
\left(\frac{a(t_0)}{a(t_{form})}\right)^{-3}\frac{\rho_{\odot}}{<\rho_M>}~,
\end{equation}
where $d^2n/dMdV_i$ is the mass spectrum modified by Hawking
evaporation until today, $a(t_0)$ and $a(t_{form})$ are the 
scale factors of the Universe nowadays and at the formation time $t_{form}$ 
(which is a function of the PBH mass), $\rho_{\odot}$ is the local halo density 
and $<\rho_M>$ is the mean matter density in the present Universe. 
The dilution factor, for $t_{form}\ll t_{eq}$, applies to all universes of interest. 
The last term 
converts the mean density into the local density under the reasonable assumption that the
clustering of PBHs follows the main dark matter component. The quantity $d^2n/dMdV_i$ 
can be obtained through the mass loss rate which reads $dM/dt=-\alpha(M)/M^2$ (by simple
integration of the Hawking spectrum multiplied by the energy of the emitted quantum)
where $\alpha(M)$ accounts for the available degrees of freedom at a given mass.
With the assumption $\alpha(M)\approx {\rm const}$ it leads to:
\be
\frac{d^2n}{dMdV_i}(M)=\frac{M^2}{(3\alpha t +
M^3)^{2/3}}\cdot\frac{d^2n_i}{dM_idV_i}
((3\alpha t + M^3)^{1/3})~.
\ee
Hence the spectrum nowadays is essentially identical to the initial one above
$M_*\equiv 3\alpha t_0\approx 5\times 10^{14}$g and proportional to $M^2$ below.

\section{Propagation and source distribution}

The propagation of the antiprotons produced by PBHs in the Galaxy
has been studied in the two zone diffusion model described in
\cite{david}, \cite{fio}.
In this model, the geometry of the Milky-Way is a cylindrical box whose
radial extension is $R=20$
kpc from the galactic center, with a disk whose thickness is $2h=200$ pc and a
diffusion halo  whose extension is still subject to 
large uncertainties.\\
The five parameters used in this model are: $K_0$,
$\delta$ (describing the diffusion coefficient $K(E)=K_0 \beta 
R^{\delta}$), the
halo half height L, the convective velocity $V_c$ and the Alfv\'en velocity
$V_a$. They have been varied within a given range determined by an exhaustive and
systematic  study of cosmic
ray nuclei data \cite{david} and chosen at their mean value. 
The same parameters used to
study the antiproton flux from a scale-free unnormalised power spectrum
in \cite{barrau4} are used again in this analysis.\\
The antiproton spectrum is affected by energy
losses when $\bar{p}$ interact with the galactic interstellar matter
and by energy  gains when reacceleration occurs.
These energy changes are described by an intricate integro--differential equation
\cite{barrau4} where a source term $q_{i}^{ter}(E)$ was added, leading to the
so-called tertiary component which corresponds to inelastic but 
non-annihilating reactions of $\bar{p}$ on interstellar matter.
Performing Bessel transforms, all the quantities can be expanded over the
orthogonal set of Bessel functions of zeroth order
and the solution of the equation for antiprotons can be explicitely obtained
\cite{david}. 
Thanks to this sophisticated model, it is no longer necessary to use 
phenomenological parameters, as in the pioneering work of MacGibbon \&
Carr \cite{McGC91}, to account for the effect of the Galactic magnetic
field. The propagation up to the Earth is naturally computed on the basis of 
well controlled and highly constrained physical processes instead of being 
described by a macroscopic parameter $\tau_{leak}$ used to enhance the local flux.

The spatial distribution of PBHs (normalized to
the local density) was assumed to follow a usual spherically symetric
isothermal profile
where the core radius $R_c$ has been fixed to 3.5 kpc and the centrogalactic distance
of the solar system $R_\odot$ to 8 kpc.
Uncertainties on $R_c$ and the consequences of a possible flatness
have been shown to be irrelevant in \cite{barrau4}.
The dark halo extends far beyond the diffusion
halo whereas its core is grossly embedded within $L$. The sources located
inside the dark matter halo but outside the
magnetic halo were shown to have a negligible contribution.

\section{Experimental data and inflationary models}

\begin{figure}
$$
\epsfxsize=9cm
\epsfysize=7cm
\epsfbox{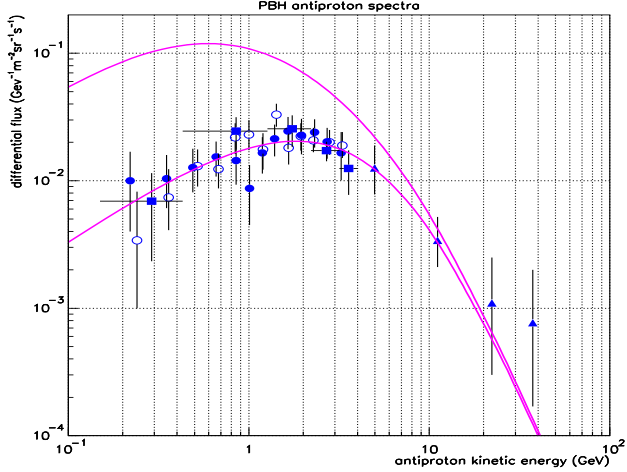}
$$
\caption{Experimental data from BESS95 (filled circles), BESS98 (circles),
CAPRICE (triangles) and AMS (squares) superimposed
with PBH and secondary spectra for $\beta(M_{peak})=5\times 10^{-28}$ (lower curve) 
and $\beta(M_{peak})=10^{-26}$(upper curves). In both cases, 
$M_{peak}=M_*$, $\epsilon=1$ and $\gamma=0.35$.}
\label{antiprotonspectrum}
\end{figure}

The astrophysical parameters decribing the propagation within the galaxy
being determined, for each set of initial parameters 
($\beta(M_{peak}), M_{peak}, \epsilon, \gamma$) defining 
the mass spectrum given in section 1, a  $\bar{p}$-spectrum is computed. 
Fig. \ref{antiprotonspectrum} gives the experimental data together with theoretical
spectra for $\beta(M_{peak})=5\times 10^{-28}$ and $\beta(M_{peak})=10^{-26}$ 
while $M_{peak}=M_*$, $\epsilon=1$ and $\gamma=0.35$. The first curve is in
agreement with data whereas the second one clearly contradicts experimental results
and excludes such a PBH density. It should be emphasized that the computed 
spectra are not only due to primary antiprotons coming from PBHs evaporation but also
to secondary antiprotons resulting from the spallation of cosmic rays on the
interstellar matter. The method used to accurately take into 
account such secondaries
is described in \cite{fio} and relies on a very detailed treatment of proton-nuclei
and nuclei-nuclei interactions near threshold thanks to a fully partonic Monte-Carlo
program. The uncertainties associated with the theoretical description of cosmic-rays
diffusion in the Galaxy (coming from degeneracy of the model with respect to several
parameters, from nuclear cross sections and from a lack of measurements of some 
astrophysical quantities) are described in \cite{fio} \& \cite{barrau4} and are 
taken into account in this work. To derive a reliable upper limit, and to account 
for asymmetric error bars in data, we define a generalized $\chi^2$ as
\bea
    \chi^2= &\sum_i&
        \frac{(\Phi^{th}(Q_i)-\Phi_i^{exp})^2}
        {(\sigma^{exp+}_i+\sigma^{th+}(Q_i))^2}\Theta(\Phi^{th}(Q_i)-\Phi^{exp}_i)\nonumber\\
        &+&\sum_i \frac{(\Phi^{th}(Q_i)-\Phi_i^{exp})^2}
        {(\sigma^{exp-}_i+\sigma^{th-}(Q_i))^2}\Theta(\Phi^{exp}_i-\Phi^{th}(Q_i))~,
\eea
where $\sigma^{th+}$ and $\sigma^{exp+}$ ($\sigma^{th-}$ and $\sigma^{exp-}$)
are the theoretical and experimental positive (negative) uncertainties,
$\Phi^{th}(Q_i)$ and $\Phi^{exp}_i$ are the theoretical and experimental antiproton
fluxes at energy $Q_i$. Requiring this $\chi^2$ to remain small enough, a
statistically significant upper limit is obtained.

\begin{figure}
$$
\epsfxsize=9cm
\epsfysize=7cm
\epsfbox{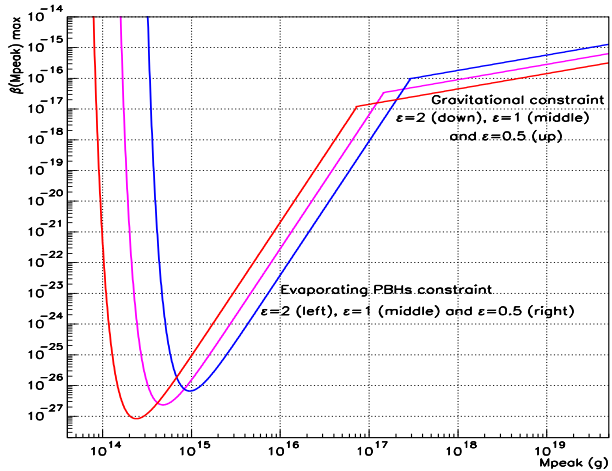}
$$
\caption{Maximum allowed value $\beta(M_{peak})$ as a function of $M_{peak}$
with $\gamma=0.35$ and $\epsilon=0.5,~1,~2$. The gravitational constraint is computed 
consistently assuming critical collapse from a bumpy mass variance at all scales.
The antiproton constraint is significantly stronger than the gravitational constraint 
in the region $M_*\lesssim M_s\lesssim 100M_*$.}
\label{limite}
\end{figure}

The maximum allowed values of $\beta(M_{peak})$ obtained
by this method are displayed in Fig. \ref{limite} as a function of $M_{peak}$ 
for $\epsilon=0.5,~1,~2$ with $\gamma=0.35$. As expected, the most
stringent limit is obtained when $M_{max}=M_*$ ({\it i.e.} $\epsilon M_{peak}= M_*$).
The curve is clearly assymetric because the mass spectrum is
exponentially supressed at $M_*$ when $M_{peak}<M_*$ whereas it decreases as a power law 
when $M_{peak}>M_*$. This constraint is significantly stronger than the
gravitational one, the requirement $\Omega_{PBH,0} < \Omega_{m,0}$,
displayed on the right hand side of the plot. 
In order to constrain inflationary models producing a bump in the mass variance, 
one has to compute the values $M_{peak}$ and $\beta(M_{peak})$. 
These will depend on the parameters of the inflationary model 
considered and can be usually traced back to the microscopic lagrangian. 
A numerical computation of $\beta(M)$ must be performed for each model 
using spectra normalized on large scales with the COBE (CMB) data for given 
cosmological background parameters, e.g. $\Omega_{\Lambda,0}=1-\Omega_{m,0}$ 
\cite{P02}. In particular, in a flat universe with $\Omega_{\Lambda,0}=0.7$, 
the mass variance at the PBH formation time is reduced by about 15\% compared 
to a flat universe with $\Omega_{m,0}=1$. 

Our results differ from those obtained in \cite{M96} in several ways. 
First, more experimental data are now available with 
much smaller errors as measurements from BESS98, CAPRICE and AMS
\cite{BESS98} where added to the first results from BESS93 \cite{Yoshimura}.
Then, a much more refined propagation model is used. This is a key point as
all the uncertainties on the astrophysical parameters used to describe the
convective, diffusive and nuclear processes occuring in the Galaxy are carefully
constrained and taken into account. The resulting antiproton flux can vary
by more than one order of magnitude between extreme acceptable astrophysical
models, making this study extremely important for the reliability of the results.
Finally, the upper limit on $\beta$ obtained in this work relies on PBH formation by 
near-critical collapse around the mass scale set by the bump in the mass variance. 
Hence, in contrast with results obtained in \cite{M96}, a constraint is obtained 
here, using eq.(\ref{mf3}), for different masses covering nearly three orders of 
magnitude. In addition, this allows us to obtain a constraint in the space of the 
inflationary free parameters for a given relevant inflationary model using the 
accurate expressions (\ref{sigF}), (\ref{TW}) in (\ref{beta}).  

To illustrate how inflationary models can be constrained, we use here a so-called 
BSI model \cite{S92} for which the quantities $M_{peak}$ and $\beta(M_{peak})$ 
can be found numerically using the analytical expression for its primordial 
power spectrum. The quantity $F(k)$ is fixed by two inflationary 
parameters $p$ and $k_s$ and exhibits a jump with large oscillations in the vicinity 
of $k_s$,
and the relative power between large and small scales is given by $p^2$ (an analytical 
expression for $F(k)$ and relevant figures can be found in \cite{S92},\cite{BBKP02}). 
This feature derives from a jump in the first derivative of the 
inflaton potential at the scale $k_s$ so that one of the slow-roll conditions is broken 
and the resulting spectrum is quite universal \cite{S92}.
Using the formalism of Section 1 one finds $k_{peak}$, which must be distinguished 
from $k_s$, as well as $\beta(M_{peak})$. Numerical calculations give $M_s\equiv 
M(t_{k_s})\approx 1.6~M_{peak}$.
We are interested in spectra with $p<1$, corresponding to more power on small scales.
In Fig. 3, the constraint on the inflationary parameter $p$ is displayed as a function of 
$M_s$. In other words each point in the plane $M_{peak},~\beta(M_{peak})$ is translated 
into 
the corresponding point $k_s,~p$. As $p$ decreases, the bump in $\sigma_H(t_k)$ and 
$\beta(M)$ increases. 
The constant spectral index $n$ (already excluded by recent CMB data) which would pass 
successfully the antiproton constraint 
corresponds to $n\approx 1.32$, only slightly less than $n$=1.33, the value satisfying the 
gravitational constraint at $M_s\simeq M_*$ \cite{BBKP02}. Indeed, as mentioned in the 
Introduction, a small change in $n$ gives a large variation in $\beta(M_*)$. 
\begin{figure}
  \begin{center} 
     \psfrag{m}[][][2.0]{$log(\frac{M_s}{g})$}
     \psfrag{p}[][][2.0]{$p \times 10^{4}$}
  \includegraphics[angle=-90,width=.75\textwidth]{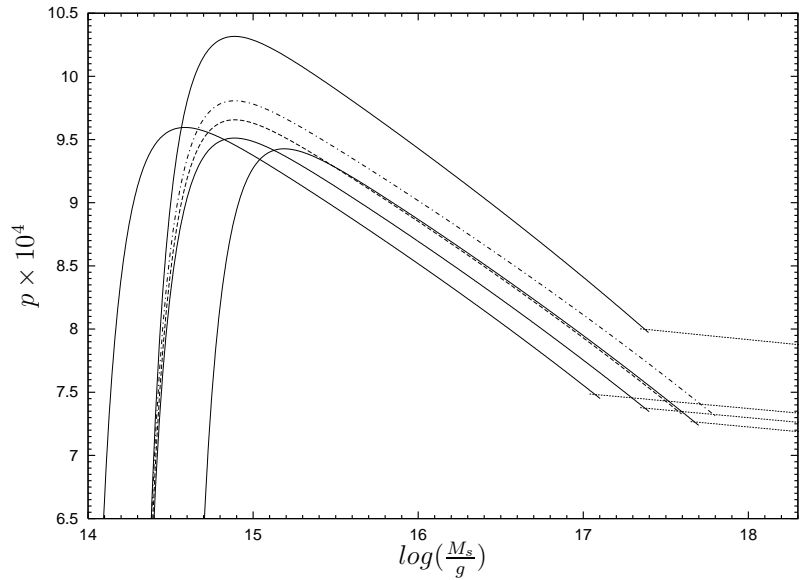}
  \end{center}
  \caption[]{The minimal value of the inflationary parameter $p$ is shown in function of 
$M_s\equiv M_H(t_{k_s})$ together with the gravitational constraint (straight lines).
For given values ($\epsilon,~\Omega_{\Lambda,0}$), the region {\it under} the 
corresponding curve is excluded by observations. 
The three solid curves at the bottom ($\epsilon=2,~1,0.5$, from the left to the right)
are the current constraints for $\Omega_{\Lambda,0}=0.7$, the upper solid curve 
corresponds to $\Omega_{m,0}=1$ and $\epsilon=1$. 
The two dashed curves, both for $(1,~0.7)$ show the improvement expected if no 
antideuteron will be found (the lower, resp. upper curve refers to AMS, resp. GAPS).} 
\label{pms}
\end{figure}

\section{Discussion}
Several improvements of our work can be expected in the forthcoming years. On
the theoretical side, a better understanding of possible QCD halos appearing
near the event horizon of PBHs should slightly alter the expected antiprotons
fluxes. The very same computation should also be performed for gamma-rays,
following, {\it e.g.} \cite{CM98}, and compared to the previously obtained 
limit on $\beta$ in \cite{CL93} and \cite{Y98}. Although essentially independent, 
the results are expected to be close to the ones obtained here.

On the experimental side, the AMS experiment \cite{Barrau5} should provide
extremely accurate data of the antiproton flux on a very wide energy
range. It should also allow to probe different solar modulation states, leading
to a better discrimination between the signal and the background \cite{Mitsui}.
Finally, it will be sensitive to low energy antideuterons which could
substantially improve the current upper limit on the PBH density. According to
\cite{Barrau6}, if no antideuteron is found in three years of data, the
limit on $\beta(M_{peak})$ will be improved by a factor of 6. Furthermore,
the GAPS project \cite{GAPS}, if actually operated in the future, would improve
the bound by a factor of 40.

\end{document}